\newif\ifprb
\prbtrue  

\ifprb
    \documentclass[%
    reprint,
    amsmath,amssymb,
    aps,
    prb,
    ]{revtex4-2}
\else
    \documentclass[%
    reprint,
    amsmath,amssymb,
    aps,
    prl,
    ]{revtex4-2}
\fi

\usepackage{graphicx}
\usepackage{dcolumn}
\usepackage{bm}
\usepackage{hyperref}
\usepackage{braket}
\usepackage{enumitem}
\usepackage{xcolor}
\usepackage{transparent}
\usepackage[export]{adjustbox}
\usepackage{tikz}
\usetikzlibrary{positioning}
\graphicspath{{figures/}}

\usepackage{caption}
\usepackage{subcaption}

\newcommand{\mat}[1]{\mathbf{#1}}

\newcommand{\rpamom}[1]{\mat{\eta}^{(#1)}}
\newcommand{\order}[1]{\mathcal{O}\left(#1\right)}

\newcommand{\toadd}[1]{{#1}}

\begin{document}

\title{Rigorous screened interactions for realistic correlated electron systems}

\author{Charles~J.~C.~Scott}
\email{charles.j.scott@kcl.ac.uk}
\affiliation{Department of Physics, King’s College London, Strand, London WC2R 2LS, United Kingdom}%

\author{George~H.~Booth}
\email{george.booth@kcl.ac.uk}
\affiliation{Department of Physics, King’s College London, Strand, London WC2R 2LS, United Kingdom}%

\date{\today}

\begin{abstract}
We derive a widely-applicable first principles approach for determining two-body, static effective interactions for low-energy Hamiltonians with quantitative accuracy. 
The algebraic construction rigorously conserves all instantaneous two-point correlation functions in a chosen model space at the level of the random phase approximation, improving upon the traditional uncontrolled static approximations. 
Applied to screened interactions within a quantum embedding framework, we demonstrate these faithfully describe the relaxation of local subspaces via downfolding high-energy physics in molecular systems, as well as enabling a systematically improvable description of the long-range plasmonic contributions in extended graphene.
\end{abstract}
\maketitle
\makeatletter

Building effective models that retain the physics of interest but strip away extraneous complexity is central to progress in understanding {physical mechanisms and emergent behaviour in} complex systems. Nowhere is this more true than interacting electron systems, where effective Hamiltonians occupy a central position in both condensed matter and quantum chemistry, from the Hubbard model to ligand field theory \cite{doi:10.1098/rspa.1964.0019,doi:10.1080/00268977500100971,10.1063/1.440050,10.1063/1.1748067}. 
A push is now underway to erode the boundary between phenomenological effective models with empirical parameters, and {\em ab initio} modelling with material specificity, 
where the act of constructing a {\em material-specific} effective Hamiltonian is increasingly the first step in a larger workflow involving its subsequent solution within a multi-method approach. This is critical to extend the scope of accurate yet computationally demanding many-body methods, placing significant urgency on approaches in which the relevant physics outside a low-energy model space is rigorously downfolded or renormalized into the effective Hamiltonian \cite{doi:10.1021/acs.jctc.0c01258,10.3389/fphy.2018.00043,chang2023learning}. In this work we propose a simple and efficient approach to effective interactions, demonstrating quantitative accuracy and specificity resulting from integrating out {\em ab initio} high-energy and long-range physics, motivated by the exact conservation of instantaneous expectation values.

A salient example in the need for accurate renormalized interactions is quantum embedding, describing correlated many-body phenomena within a local subspace \cite{RevModPhys.90.025003,RevModPhys.78.865,doi:10.1021/acs.accounts.6b00356,ditte2023molecules}. 
The missing interactions with states outside the subspace should renormalize the effective subspace interactions in real materials, generally necessitating the use of effective interactions that are often parameterized empirically via Hubbard or Hund terms \cite{PhysRevB.86.085117,PhysRevB.83.121101,PhysRevB.57.4364}, or downfolded from other theories \cite{ditte2023molecules, PhysRevB.92.045113, PhysRevB.103.125130,PhysRevLett.102.176402}. In quantum chemistry, analogous subspaces are often described by a `complete active space' (CAS), whereby a small number of low-energy mean-field orbitals are chosen for an accurate treatment of the strong correlation in `multi-reference' approaches \cite{doi:10.1021/acs.chemrev.9b00496}. More broadly, a wide range of both qualitative and quantitative studies into correlated materials rely on first obtaining appropriately screened effective interactions of a simplified model from an {\em ab initio} starting point \cite{alavi20,doi:10.1021/acs.jctc.0c01258}.

It is often argued that the random phase approximation (RPA) contains much of the physics missing from low-energy models, and is the appropriate theory in which to construct effective interactions \cite{doi:10.1146/annurev-physchem-040215-112308,ditte2023molecules,PhysRevB.83.121101,10.1063/5.0011195}. As an infinite resummation of the bubble diagrams, it correctly describes the long-range charge fluctuations, plasmons, high-energy scattering and many-body dispersion that predominates in high-density metallic and semi-conducting extended systems \cite{PhysRev.92.609,PhysRev.106.364,PhysRevB.75.235102}. 
While its traditional formulation lacks exchange or ladder diagrams, their contributions generally decay more rapidly and can often be captured within the model space, with approaches to screen beyond RPA still an active research area \cite{PhysRevB.104.045134,PhysRevB.92.045113,PhysRevB.103.125130}.

The {\em constrained} RPA (cRPA) method has therefore become a widespread choice for deriving low-energy effective interactions from first principles, applied from molecules to Mott insulators and routinely as a precursor to quantum embedding methods \cite{PhysRevB.70.195104, PhysRevB.77.085122,PhysRevB.83.121101,PhysRevB.104.045134, PhysRevB.91.245156,PhysRevB.103.125130, PhysRevB.86.165105,Hausoel2017, PhysRevB.87.165118, Loon18, PhysRevB.91.125142, Jang2016, PhysRevB.91.241111, PhysRevB.100.205138, PhysRevLett.107.266404, PhysRevB.85.045132, yoshimi2022comprehensive,RevModPhys.90.025003,PhysRevB.104.245114,doi:10.1021/acs.jctc.0c01258,doi:10.1021/acs.jctc.2c00240}. Since the RPA is a well-defined diagrammatic theory, the bubbles corresponding to the polarizability {\em within} the model space ($\Pi_m(\omega)$) can be removed from the total polarization ($\Pi_{\rm ext}(\omega)=\Pi(\omega)-\Pi_m(\omega)$). The screened interactions, $\mathcal{U}(\omega)$, are then found from the infinite RPA resummation coupling this space to the external degrees of freedom via this scattering channel, as
\begin{align}
    \mathcal{U}(\omega) = v+v\Pi_{\rm ext}(\omega)v+v\Pi_{\rm ext}(\omega)v\Pi_{\rm ext}(\omega)v + \dots , \label{eq:screening}
\end{align}
where $v$ denotes the bare Coulomb interaction.
In this way, double counting of correlated effects are avoided once the resulting effective model is solved, yet direct scattering events to all orders between the two spaces are included \cite{PhysRevB.70.195104}. The cRPA approach is also internally consistent: an RPA calculation on the resulting effective Hamiltonian with interactions $\mathcal{U}(\omega)$, will give exactly the results of the RPA on the full system. More specifically, the cRPA ensures that the RPA density-density (dd) response function, $\chi(\omega)$, of the model space with effective interactions $\mathcal{U}(\omega)$, is identical to the projection of the RPA full system dd-response into the model space. In this way, the effective interactions can also be seen to be `state universal', providing the correctly renormalized interactions for the entire RPA spectrum of the model subspace.

While this puts the cRPA approach on a solid footing, few methods are computationally tractable with the resulting {\em dynamical} interactions of Eq.~\ref{eq:screening}. Practical necessity therefore forces the widespread approximation of taking the static, $\omega \rightarrow 0$ limit of the screened interaction of Eq.~\ref{eq:screening} (static-cRPA). This uncontrolled approximation is qualitatively justified in capturing the relevant long-wavelength behaviour where the RPA is accurate \cite{PhysRevB.104.045134}, however forces us to entirely give up on rigorous conservation of any expectation values from the RPA. There is significant scepticism of its accuracy, with the cRPA interactions often over-screening the physics \cite{PhysRevB.104.045134, PhysRevB.91.245156, PhysRevB.98.235151, PhysRevB.92.045113, PhysRevB.103.125130}. Far from purely a small quantitative shift, this can result in wrong energy ordering of states and phases in the resulting model, or missing spectral weight transfer to plasmon satellites \cite{PhysRevLett.109.126408,PhysRevB.57.4364,PhysRevB.91.241111}.

A further serious technical limitation of the cRPA approach is that the model space must be chosen from a selection of low-energy bands, rather than directly as local degrees of freedom. Specifically, it requires the irreducible polarizability (defined by the reference mean-field) to have no coupling between excitations within the model space and ones in the rest of the system, otherwise $\mathcal{U}(\omega)$ alone is not sufficient to reproduce $\chi(\omega)$.
Effective local interactions are therefore found via localization {\em after} determining the screening of these low-energy bands. However, this is problematic, missing out on screening within the low-energy bands themselves if the resulting local interactions are subsequently truncated (e.g. to Hubbard/Hund form), and causing further difficulties when it is not possible to fully represent the relevant subspace (e.g. atomic $d$-shells) in the low energy bands due to significant hybridization with other states\cite{PhysRevB.80.155134, PhysRevB.86.165105}. In addition, the resulting cRPA screened interactions modify the reference mean-field state, resulting in unintended changes to the subspace density and bandstructure \cite{PhysRevB.91.245156,PhysRevB.98.235151,AryasetiawanNotes}. 

To motivate our approach, we ask an alternative pertinent question: What physical quantities {\em can} we rigorously conserve at the RPA level in a chosen model space, under the constraint that the resulting effective interactions must remain {\em static} and {\em two-body}? This leads to the development of the `moment-constrained' RPA (mRPA) approach, which exactly conserves the {\em instantaneous} part of the two-point dd-response in the model subspace, as well as the reference state. We argue that conserving physical expectation values in the construction of effective static interactions provides a more rigorous foundation than the widespread static-cRPA approximations. 

\emph{Moment-constrained RPA:-} The RPA can be formulated 
as a quasi-bosonic eigenvalue problem in the space of particle-hole excitations and de-excitations of a reference state \cite{casida1995time}
,
\begin{align} \label{eq:Casida-eq}
\begin{bmatrix}
    \mat{A} & \mat{B} & \\
    \mat{-B} & \mat{-A} & 
\end{bmatrix} \begin{bmatrix} \mat{X} & \mat{Y} \\ \mat{Y} & \mat{X} \end{bmatrix} = 
\begin{bmatrix}
\mat{X} & \mat{Y} \\ \mat{Y} & \mat{X}
\end{bmatrix} \begin{bmatrix} \mat{\Omega} & \mat{0} \\ \mat{0} & -\mat{\Omega} \end{bmatrix} ,
\end{align}
with all blocks being of dimension given by the product of the number of hole and particle states. We define $\mat{A}=\mat{\Delta} + v$, where $\mat{\Delta}$ is a diagonal matrix of particle-hole excitation energies (provided by a mean-field), defining the poles of $\Pi(\omega)$, with $\mat{B}=v$ providing the coupling between the excitations and de-excitations via the bare Coulomb interaction, $v$, in the particle-hole ($ph$) channel. The diagonal matrix $\mat{\Omega}$ provides the poles of the RPA dd-response function, $\chi(\omega)$, with the residues defined by the amplitudes of the excitations and de-excitations, $\mat{X}$ and $\mat{Y}$ respectively. These form a biorthogonal set of eigenvectors, providing the relations $(\mat{X}+\mat{Y})^{-1}=(\mat{X}-\mat{Y})^T$ and $(\mat{X}-\mat{Y})^{-1}=(\mat{X}+\mat{Y})^T$. We can expand $\chi(\omega)$ as a Laurent series, with its dynamics fully characterized by the moments of its spectral distribution \cite{PhysRevB.104.245114,10.1063/5.0143291}, as
\begin{equation} \label{eq:def_eta}
    \eta^{(n)} = (\mat{X}+\mat{Y}) \mat{\Omega}^n (\mat{X}+\mat{Y})^T  \quad; \quad n \in {\mathbb Z}. 
\end{equation}

The zeroth moment of the distribution, $\eta^{(0)}$, characterizes the instantaneous part of the correlated dd-response, as $\langle ({\hat c}^\dagger_i {\hat c}_a+{\hat c}_a^\dagger {\hat c}_i) ({\hat c}_j^\dagger {\hat c}_b+{\hat c}_b^\dagger {\hat c}_j)\rangle - \langle {\hat c}^\dagger_i {\hat c}_a+{\hat c}_a^\dagger {\hat c}_i \rangle\langle {\hat c}_j^\dagger {\hat c}_b+{\hat c}_b^\dagger {\hat c}_j\rangle$, 
summed over the same-spin particle-hole (de)excitations, denoted by indices $(a,i)$ and $(b,j)$. This describes the correlated contribution to the two-body reduced density matrix, and all resulting static expectation values at the RPA level~\cite{toulouse2009adiabatic, toulouse2010range, angyan2011correlation}. It is this quantity (through to first order) which we aim to rigorously conserve within a chosen model space with our effective interactions. 
Crucially, this can be achieved while maintaining {\em static} renormalized effective interactions in the model space, preserving symmetries, and without changing the reference mean-field in the model space.

The structure of the RPA equations imposes a relation between the first two dd-moments \cite{10.1063/5.0143291},
\begin{align}
    \eta^{(1)} &= (\mat{A}-\mat{B}) = \eta^{(0)} (\mat{X}-\mat{Y}) \mat{\Omega} (\mat{X}-\mat{Y})^T \eta^{(0)} \\
    &= \eta^{(0)} (\mat{A}+\mat{B}) \eta^{(0)} . \label{eq:mom_relations}
\end{align}
Inserting the definitions of $\mat{A}$ and $\mat{B}$, we find an equation linear in the interaction. This can be analytically inverted to find an interaction kernel that under the RPA gives rise to a desired $\eta^{(0)}$ and $\eta^{(1)}$,
\begin{equation}
    v = \frac{1}{2} \left( (\eta^{(0)})^{-1} \eta^{(1)} (\eta^{(0)})^{-1} - \mat{\Delta} \right) . \label{eq:inv_v}
\end{equation}
Therefore, by substituting all quantities on the RHS of Eq.~\ref{eq:inv_v} for their projection of the full RPA into the chosen model space, an effective static model space interaction can be found, $\mathcal{U}_{\textrm{mRPA}}$. This ensures that the model space RPA with $v \rightarrow \mathcal{U}_{\textrm{mRPA}}$ rigorously conserves all full system RPA expectation values derived from $\rpamom{0}$ and $\rpamom{1}$ by construction, which includes all instantaneous correlators in the model space. A similar approach to conserve higher-order moments results in equations non-linear in the resulting interaction, and subsequently necessitates a dynamical component to the resulting interaction, as expected to recover the dynamical cRPA limit of the full dd-response \cite{PhysRevB.104.245114}. Focusing on conservation of just the first two dd-moments enables a fully static model space screened interaction.

The model space with $\mathcal{U}_{\textrm{mRPA}}$ can also reproduce the subspace contribution to the RPA correlation energy due to its dependence on $\rpamom{0}$, as $E_{\mathrm{corr}}^{\mathrm{RPA}} = \frac{1}{2} \mathrm{Tr}[\mat{\eta}^{(0)} (\mat{A} + \mat{B}) - \mat{A}]$, exploited in 
\ifprb
App.~\ref{app:graphene} 
\else
the SI (Sec. IV)
\fi
to compute energetic corrections to the model space. 
Beyond this, the first two spectral moments of the $GW$ self-energy in the subspace are also exactly described with mRPA interactions, indicating a formal conservation of certain correlated one-body properties \cite{10.1063/5.0143291}. Symmetries, including spin-independence of the resulting effective interactions are also exactly preserved, derived in
\ifprb
App.~\ref{app:spin-independence}. 
\else
the SI (Sec. III).
\fi
These rigorous properties provide a robust footing for use of $\mathcal{U}_\textrm{mRPA}$ in subsequent correlated treatments. 

Building $\mathcal{U}_\textrm{mRPA}$ via Eq.~\ref{eq:inv_v} requires the component of $\rpamom{0}$ and $\mathbf{\Delta}$ in the subspace. Construction directly from Eqs.~\ref{eq:Casida-eq}-\ref{eq:def_eta} entails a prohibitive $\order{N^6}$ scaling, but this is reduced to an efficient $\order{N^4}$ following the approach
\ifprb
of App.~\ref{app:compute}.
\else
detailed in the SI (Sec. I).
\fi
\toadd{The only constraint on the choice of subspace that the cluster excitation space is an orthogonal projection of the full system excitation space, i.e. that particle- and hole-like character of the subspace orbitals as defined by the reference state is preserved.
While this is trivially true where cRPA is valid via a subspace selected from mean-field bands, this is a far looser requirement allowing for a subspace where a non-diagonal $\Pi(\omega)$ couples the subspace to its environment to be considered. This admits direct mRPA screening of arbitrary (e.g. local atomic-like) subspaces by at most doubling their size by conserving the reference mean-field density matrix over the subspace\cite{PhysRevLett.109.186404,sekaran2023unified}, or local spaces formed by localizing hole and virtual bands separately before screening. This direct screening of local subspaces therefore includes screening via long-range low-energy band transitions precluded in traditional cRPA, enabling direct application to {\em ab initio} quantum embedding clusters.}

Interestingly, the resulting $\mathcal{U}_{\textrm{mRPA}}$ only screens interactions in the $ph$ channel \toadd{(which is expected to be the dominant long-range contribution to screening)}, rather than the full four-point interaction. 
Other approximations (e.g. $T$-matrix for $pp$ diagrams~\cite{BICKERS1989206,PhysRevA.88.030501,PhysRevB.91.235114,PhysRevLett.80.2389}) would screen other interaction channels in an analogous formulation\toadd{, and future work can consider the effect on mRPA interactions from these other channels}. It is justified that only the $ph$ channel interaction is screened \toadd{in mRPA unlike cRPA, since RPA itself is fully determined by this component of the interaction, and mRPA is constructed to describe this RPA physics as opposed to the use of the screening equation (Eq.~\ref{eq:screening}) in cRPA. 
This feature of only screening the $ph$ interactions}, along with the conservation of the reference density, ensures that the occupied reference bandstructure (Hartree--Fock or Kohn--Sham) is unchanged with mRPA screened interactions. 

\begin{figure}
    \centering
    \includegraphics[width=\linewidth]{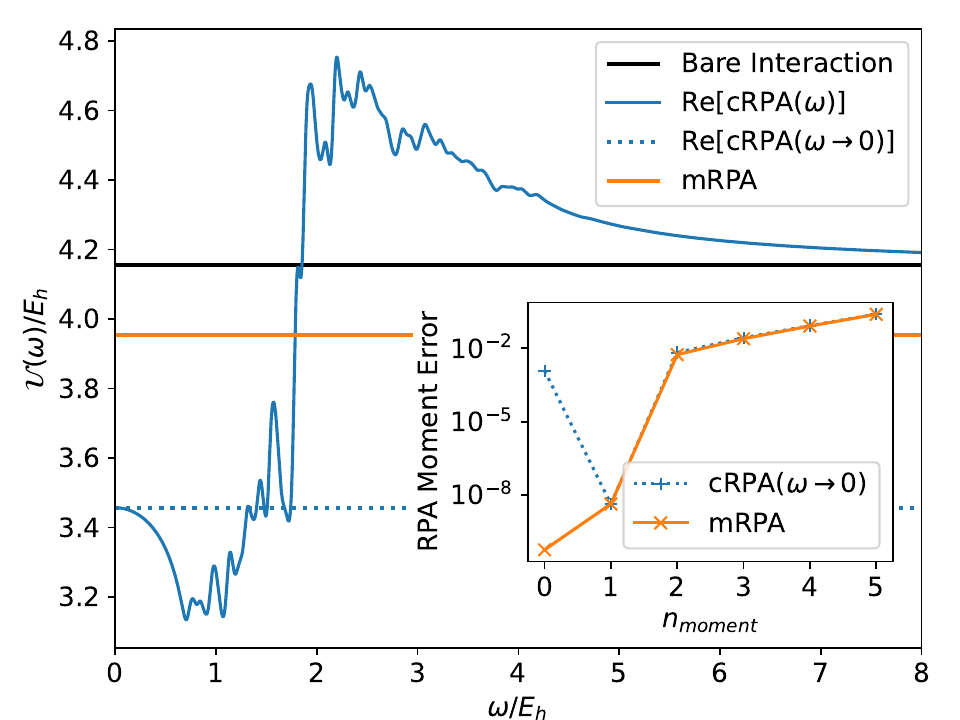}
    \caption{Trace of the screened and bare interaction, $\mathcal{U}_{pq,pq}$ in a (10,10) CAS space of Benzene in a cc-pVDZ basis. Shown are the dynamical cRPA interactions, as well as the widely used static limit, and the static-by-construction mRPA interactions. Inset: Error in the RPA moments as an expansion of the dynamical subspace dd-response. The first two mRPA dd-moments are exact by construction, yet also exhibit marginally smaller errors for higher moments compared to the static cRPA interaction. Moment error is computed as the mean squared error of $\rpamom{n}$ over the subspace, normalized by the exact $||\rpamom{n}||_2$ at each order.}
    \label{fig:benzene_cRPA}
\end{figure}

\emph{Results:-} We begin by benchmarking the screening of the low-energy Hartree--Fock orbitals of Benzene (an active space of five hole and five particle states). Benzene has previously been considered a paradigmatic example in this context \cite{PhysRevB.104.045134}, while also small enough to enable comparison to high-level reference results. In Fig.~\ref{fig:benzene_cRPA} we show the bare Coulomb, dynamical and static-cRPA, and mRPA subspace interactions traced over all channels. 
The $\mathcal{U}_{\textrm{mRPA}}$ is generally less screened than static-cRPA, noting the recent evidence that static-cRPA overscreens interactions \cite{PhysRevB.91.245156, PhysRevB.98.235151, PhysRevB.92.045113}.
We can consider the accuracy of resulting subspace observables by comparing to coupled-cluster (CCSD)\cite{https://doi.org/10.1002/wcms.1340,10.1063/5.0006074,RevModPhys.79.291}. CCSD encodes the wavefunction in $T$-amplitudes, and we can consider the projection of the full-system $T$-amplitudes into the active space as an accurate subspace description. \toadd{This is used to benchmark {\em subspace-only} CCSD calculations} with the different static interactions \footnote{Coupled-cluster is defined in terms of the Fock matrix and fluctuation potential. The virtual-virtual block of the Fock matrix is shifted by $\Delta f_{ab} = \sum_{i \in \textrm{occ}} (ia|bi) - \mathcal{U}_{ia,ib}$ due to the screened $ph$ mRPA interactions, but other blocks are unchanged ensuring the reference-independence of mRPA.}. 
We find the mean squared error of the subspace CCSD $T_2$-amplitudes to be 1.10 for bare subspace interactions, reduced to 0.25 for static-cRPA and only 0.20 for mRPA, indicating that {\em all} ground-state expectation values within the subspace can expected to be more faithfully reproduced with mRPA interactions than the alternatives. 

While we expect ground states to be particularly faithful given the mRPA conservation of instantaneous correlators, we can also compare the ability of these effective interactions to reproduce the full subspace excitation spectrum. Fig.~\ref{fig:benzene_cRPA} quantifies this via the errors in the subspace RPA moment expansion of Eq.~\ref{eq:def_eta} which fully characterizes the dd-response. While the first two moments are exact for mRPA by construction (and $\rpamom{1}$ for static-cRPA), the errors in higher moments are also marginally reduced in their relative error compared to static-cRPA, indicating that the fidelity of the subspace dd-response over all frequencies is at least as accurate as static-cRPA.


We now consider the diverse W4-11 test set of 150 molecules, exhibiting a wide range of bonding, radical and correlated physics \cite{KARTON2011165}. 
Moving towards direct application of mRPA to quantum embedding methodologies, we consider fragmenting each molecule into individual atoms comprising a minimal set of their intrinsic valence atomic orbitals (IAOs)\cite{doi:10.1021/ct400687b}. These IAO fragments are augmented with an interacting bath of at most the same dimension as the IAO fragment, following the static density matrix embedding (DMET) approach\cite{PhysRevLett.109.186404,doi:10.1021/ct301044e,sekaran2023unified,doi:10.1021/acs.jctc.6b00316,doi:10.1021/acs.jctc.9b00933,PhysRevX.12.011046}. This defines multiple small atomic subspaces for each molecule, which can be individually solved via CCSD with either bare or mRPA interactions. The resulting subspace CCSD states are recombined to provide a total energy estimator over the subspaces for each molecule (see Refs.~\onlinecite{PhysRevX.12.011046, doi:10.1021/acs.jctc.2c01063} for more details). These are compared to the energy of the exact full-system CCSD projected into these subspaces. This discrepancy quantifies the ability of the mRPA to account for relaxation of these embedded atomic descriptions due to the neglected interactions with the rest of the molecule. 

\begin{figure}
    \centering
    \includegraphics[width=\linewidth]{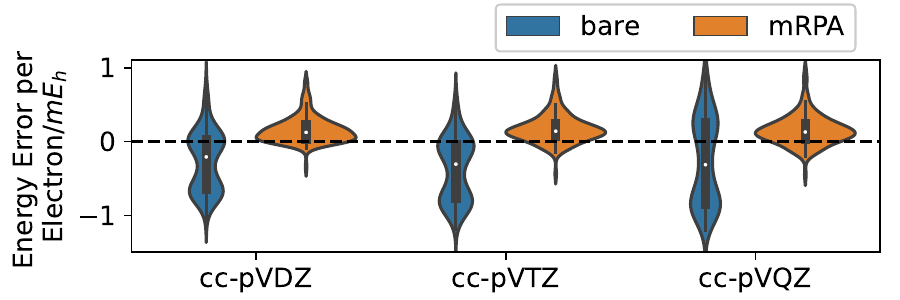}
    \caption{Distribution of energy errors in atomically fragmented DMET-CCSD with bare or mRPA interactions across the W4-11 test set of 150 molecules, in increasingly large basis sets. Errors are computed comparing to the full-system CCSD of each molecule projected into each subspace.}
    \label{fig:W11}
\end{figure}

Figure~\ref{fig:W11} shows this error aggregated over the test set in increasingly large basis sets, where screening of these low-energy subspaces by higher-energy scattering dominates. The mRPA interactions screen the fragments defined by the embedding, resulting in a substantial $\sim66\%$ reduction in both the \toadd{mean absolute error and standard deviation of the energy error} across the data set. This consistent reduction in error across larger basis sets points for these diverse systems attests to the broad applicability of the mRPA interactions for correlated systems. We note that cRPA cannot be easily compared in this context, due to the difficulties in directly screening subspaces that are not mean-field orbitals, as discussed previously.


Finally, we consider the application of mRPA to extended systems where long-range collective plasmons strongly renormalize local properties, which are difficult to describe in a local subspace\cite{PhysRevLett.106.236805,PhysRevB.104.045134}. Furthermore, we demonstrate systematic improvability of this screened local subspace. This is achieved in a consistent framework via extending the bath space, formally including additional states which exactly minimize the error of the subspace RPA $\rpamom{0}$ with bare interactions, thereby spanning physics at longer length-scales in the model subspaces.
The algebraic construction of these additional bath states relies on evaluation of the same quantities as the mRPA interactions, and systematically enlarges the local correlated subspaces in an optimal way to completeness, with details in
\ifprb
App.~\ref{app:RPAbath}.
\else
the SI (Sec. III).
\fi
This approach is inspired by the unscreened perturbative bath expansion of Ref.~\onlinecite{PhysRevX.12.011046}, but here adapted for a screened embedding.

\begin{figure}
    \centering
    \includegraphics[width=\linewidth]{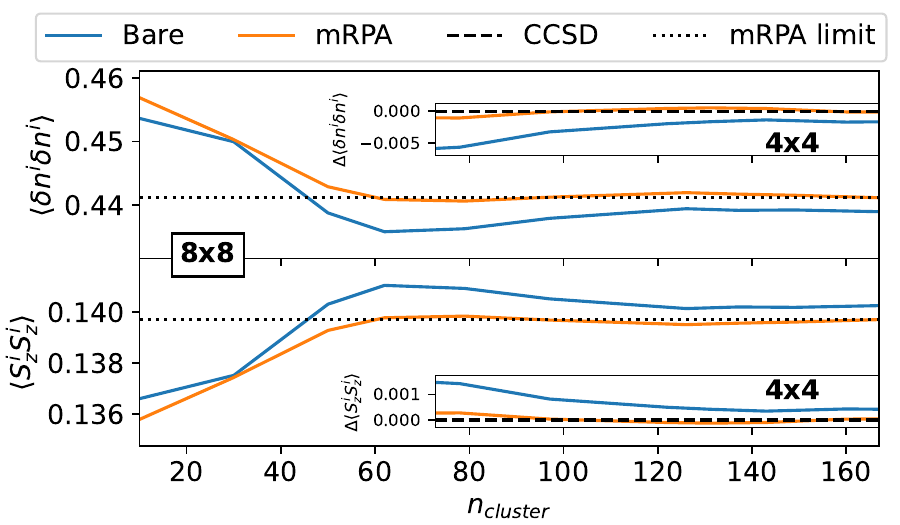}
    \caption{Convergence of the CCSD local $2p_z$-orbital charge (top) and spin (bottom) two-body fluctuations in $8\times8$ Graphene sheets with increasing size of embedded cluster, for both bare Coulomb and mRPA cluster interactions. Inset: Error in these two-body correlators in $4\times4$ $\textbf{k}-$point meshes, where comparison to exact CCSD is possible.}
    \label{fig:graphene-main}
\end{figure}

We converge semi-metallic graphene sheets with a fully {\em ab initio} CCSD description, embedding atomic fragments and systematically enlarging the \toadd{bath space for each fragment. On enlarging, the bath interactions provide increased screening of the fragments explicitly, and the mRPA static screening is correspondingly reduced in a fashion that precludes double-counting of the fragment screening and naturally converges to an in-method (CCSD) exact limit}. Figure~\ref{fig:graphene-main} shows fully {\em ab initio} two-body instantaneous charge and spin fluctuations in the local atomic $2p_z$ space, on top of a symmetry-preserving Hartree--Fock reference \cite{10.1063/5.0055191,doi:10.1021/acs.jctc.2c01063}. The correlated treatment quantitatively changes these two-body correlators, but bare subspace interactions overestimate the magnitude of the changes from mean-field. Importantly, the bias in these correlators with the bare interactions appears unable to be compensated for with increasing bath, indicating the importance of coupling to truly long-range $\mathbf{q}\rightarrow0$ plasmonic modes for an appropriate relaxation of these local properties, impractical in a local embedding with bare interactions.

In contrast, the mRPA screened interactions fold in this coupling, resulting in a rapid and stable convergence with increasing bath size, corroborated with smaller $4\times4$ $\mathbf{k}$-point meshes where comparison to the full system CCSD result is possible and the same qualitative behaviour is observed. While $\mathcal{U}_{\mathrm{mRPA}}$ is static, it nevertheless integrates over all RPA diagrams, including plasmonic contributions required to appropriately relax the subspace. This even impacts on the convergence of magnetic correlators in the subspace, despite not being described in the $\rpamom{0}$ description of RPA density fluctuations. The results indicate that the plasmonic coupling therefore suppresses the tendency towards formation of atomic magnetic moments in graphene. In 
\ifprb
App.~\ref{app:graphene} 
\else
the SI (Sec. IV)
\fi
we also demonstrate improvements in {\em non-local} magnetic fluctuations as well as energetics in this system, and the insensitivity of these results to choice of underlying reference mean-field theory.

In summary, the developed `mRPA' efficiently integrates over all external RPA diagrams to provide a manifestly static effective interaction for low energy or local models, conserving all instantaneous RPA correlators in a chosen subspace by construction. For multi-method workflows, we show this provides a systematic and principled route for the inclusion of long-range and high-energy screening in a local correlation framework. This also opens opportunities in the push for fully improvable and non-empirical quantum embedding, with reliable convergence in extended systems coupled to long-range coherent quasiparticles.

\ifprb
\begin{acknowledgments}
    The authors gratefully acknowledge support from the European Union’s Horizon 2020 research and innovation programme under grant agreement No. 759063. We are grateful to the UK Materials and Molecular Modelling Hub for computational resources, which is partially funded by EPSRC (EP/P020194/1 and EP/T022213/1). All calculations were performed and can be reproduced with the {\tt Vayesta} quantum embedding code~\footnote{{\tt Vayesta} code available at {\tt https://github.com/BoothGroup/Vayesta}.}.
\end{acknowledgments}
\fi

%

\ifprb
\appendix
\section{Efficient computation of RPA subspace moments} \label{app:compute}

We consider an efficient approach for the construction of the subspace projection of the RPA quantities $\rpamom{0}$ and $\rpamom{1}$. These are defined by their relation in Eq.~\ref{eq:mom_relations}, as required for the subsequent computation of the subspace screened interaction $\mathcal{U}_{\mathrm{mRPA}}$ via Eq.~\ref{eq:inv_v}. This is similar to derivations found in Ref.~\onlinecite{10.1063/5.0143291} in the context of efficient $GW$ theory, which itself is based on efficient RPA correlation energy computation of Ref.~\onlinecite{Furche2010}. Ultimately, since the working equations of mRPA below are algorithmically similar to many RPA and $GW$ methods, we expect the computational feasibility of efficient mRPA implementations to be similar. Starting from Eq.~\ref{eq:mom_relations}, we can multiply both sides from the right by $(\mat{A}+\mat{B})$ to give
\begin{equation}
    (\mat{A}-\mat{B})(\mat{A}+\mat{B}) = [\rpamom{0}(\mat{A}+\mat{B})]^2 . \label{eq:rpamom0}
\end{equation}
We then employ a standard low-rank decomposition of the two-electron repulsion integrals (e.g. via density fitting or Cholesky decomposition) in the particle-hole channel, as
\begin{equation}
v_{ia,jb} \simeq \sum_P^{N_{\mathrm{aux}}} V_{ia,P} V_{jb,P} = \mat{V}\mat{V}^T , \label{eq:RI_eri}
\end{equation}
where we use $P,Q,\dots$ to index elements of this auxiliary (RI) basis, whose dimension $N_{\mathrm{aux}}$ scales $\order{N}$ with system size, and $i,j,\dots$ ($a,b,\dots$) denote hole (particle) spin-orbital energies from the mean-field reference description (note App.~\ref{app:spin-independence} details direct construction of the spin-integrated mRPA interaction). Writing $(\mat{A}\pm \mat{B})$ explicitly in the (direct) RPA form, as
\begin{align}
    (\mat{A}-\mat{B})_{ia,jb} &= (\epsilon_a - \epsilon_i) \delta_{ij} \delta_{ab} = \mat{\Delta} , \label{eq:AminB} \\
    (\mat{A}+\mat{B})_{ia,jb} &= (\epsilon_a - \epsilon_i) \delta_{ij} \delta_{ab} + 2 \sum_P V_{ia,P} V_{jb,P} \nonumber \\
    &= \mat{\Delta} + 2 \mat{V} \mat{V}^T, \label{eq:AplusB}
\end{align}
we can cast Eq.~\ref{eq:rpamom0} as
\begin{align}
    \rpamom{0} &= [(\mat{A}-\mat{B})(\mat{A}+\mat{B})]^{1/2} (\mat{A}+\mat{B})^{-1} \\
    &= (\mat{\Delta}^2 + 2 \mat{\Delta} \mat{VV}^T)^{1/2} (\mat{\Delta}+2 \mat{V} \mat{V}^T)^{-1}.
\end{align}
Using the Woodbury matrix identity, we can rewrite the first part of this term as
\begin{align}
\mat{T} &= (\mat{\Delta}+2\mat{V}\mat{V}^T)^{-1}  \\
&= \mat{\Delta}^{-1} - 2\mat{\Delta}^{-1} \mat{V}(\mat{I}+2\mat{V}^T \mat{\Delta}^{-1} \mat{V})^{-1} \mat{V}^T \mat{\Delta}^{-1} . \label{eq:Woodbury}
\end{align}
This now only requires the inversion of the diagonal matrix, $\mat{\Delta}$, and a matrix of dimension $N_{\mathrm{aux}}$, with the overall $ov \times N_{\mathrm{aux}}$ matrix able to be constructed in $\mathcal{O}[N_{\mathrm{aux}}^3 + N_{\mathrm{aux}}^2ov]$ time, where $o$ and $v$ are the number of hole and particle states respectively.
The second part of the expression can be computed efficiently using the definition of the matrix square-root as an integration in the complex plane \cite{matrixsqrt},
\begin{equation}
\mat{M}^{\frac{1}{2}}=\frac{1}{\pi} \int_{-\infty}^{\infty} \left( \mat{I}-z^2(\mat{M}+z^2 \mat{I})^{-1} \right) dz \label{eq:matsqrt} .
\end{equation}
This results in
\begin{align}
\mat{\eta}^{(0)} &= (\mat{\Delta}^2 + 2 \mat{\Delta V} \mat{V}^T)^{\frac{1}{2}} \mat{T}  \\
&= \frac{1}{\pi}\int_{-\infty}^{\infty} \left( \mat{I} - z^2 (\mat{\Delta}^2 + 2\mat{\Delta V V}^T + z^2 \mat{I})^{-1} \right) dz \mat{T} . \label{eq:NI}
\end{align}
We can apply the Woodbury matrix identity a second time to this integrand, and simplify the notation by introducing the intermediates,
\begin{align}
\mat{F}(z) &= (\mat{\Delta}^2 + z^2 \mat{I})^{-1} \label{eq:F} \\
\mat{Q}(z) &= 2 \mat{V}^T \mat{F}(z) \mat{\Delta V} , \label{eq:Q}
\end{align}
where $\mat{F}(z)$ is a diagonal matrix in the $ov$ space, and $\mat{Q}(z)$ is a $N_{\mathrm{aux}}\times N_{\mathrm{aux}}$ matrix which can be constructed in $\mathcal{O}[N_{\mathrm{aux}}^2 o v]$ time (the leading-order computational step). 
This casts Eq.~\ref{eq:NI} into the form
\begin{align}
\eta^{(0)} = \frac{1}{\pi}& \int_{-\infty}^{\infty} \left[ \mat{I} - z^2 \mat{F}(z) +\right. \label{eq:NI_efficient}\\
& \left. 2 z^2 \mat{F}(z)\mat{\Delta V}(\mat{I}+\mat{Q}(z))^{-1}\mat{V}^T \mat{F}(z) \right] dz  \mat{T}. \nonumber
\end{align}

This integrand bears much similarity to quantities computed in efficient RPA implementations, or Matsubara $GW$ implementations \cite{PhysRevB.106.235104,doi:10.1021/acs.jctc.0c00704,Furche2010}, where $z$ represents an imaginary frequency variable.
Following Ref.~\onlinecite{10.1063/5.0143291}, it is possible to analytically integrate the mean-field contribution to this quantity, as well as partially integrate the slowest order decaying contribution to the integrand in $z$, to improve the efficiency of numerical quadrature. This results in the expression
\begin{align}
\eta^{(0)} &= \mat{\Delta} \mat{T} \nonumber \\
&+ 2 \int_0^{\infty}  e^{-t \mat{\Delta}} \mat{\Delta V} \mat{V}^T e^{-t \mat{\Delta}} \mat{T} dt  \nonumber \\
&+ \frac{2}{\pi}\int_{-\infty}^{\infty} z^2 \mat{F}(z) \mat{\Delta V}\left((\mat{I}+\mat{Q}(z))^{-1}-\mat{I} \right) \mat{V}^T \mat{F}(z) \mat{T} dz . \label{eq:final_NI_zero_mom_main}
\end{align}
The first numerical integral in Eq.~\ref{eq:final_NI_zero_mom_main} (where the integrand decays exponentially) is computed via Gauss-Laguerre quadrature, while the second (where the integrand decays as $\mathcal{O}[z^{-4}]$) is evaluated via Clenshaw--Curtis quadrature. Quadrature points and weights are optimized according to the protocol in Ref.~\onlinecite{10.1063/5.0143291}, resulting in exponential convergence with number of points, and a practical necessity for no more than $24$ points.

While this describes the calculation of the full system $\rpamom{0}$, only the component of $\rpamom{0}$ in the subspace of excitations is required. The projector from the space of full system particle-hole excitations into this subspace is constructed as 
\begin{equation}
    Z_{ia,pq} = C_{pi} C_{qa} = \mat{Z} ,
\end{equation}
where $C_{pi}$ are the coefficients of the model space in the basis of the hole and particle reference states. Including $\mat{Z}$ explicitly on both sides of Eq.~\ref{eq:final_NI_zero_mom_main} and optimizing the contraction order of all quantities, it can be shown that the scaling for the construction of the subspace component is $\order{N^4}$, with the rate limiting step the $\order{N_{\textrm{aux}} ov}$ construction of the $\mat{Q}(z)$ matrix. This can be reduced to $\order{N^3}$ with the introduction of doubly-factorized integrals such as those from tensor hypercontraction or forms found in the space-time approach to $GW$ theory, or from an iterative approach to the direct construction and inversion of the $(\mat{I}+\mat{Q}(z))$ matrix with local projectors, which will be considered in future work. 
Furthermore, since the $\rpamom{1}$ quantity is diagonal, finding the subspace component of this quantity is trivial. 

Finally, we note that this $\order{N^4}$ formal scaling is unchanged if you want to find the mRPA screened interactions of a single subspace (even one whose size grows linearly with system size) or multiple local subspaces simultaneously, since the full system $\mat{Q}(z)$ matrix construction and inversion is still performed once, and other contractions do not scale worse than this. This feature is exploited in the main text in the results of Fig.~\ref{fig:W11}, where each atom is considered part of its own screened subspace interactions, and thus the number of subspaces which you need to find the screened interactions for increases linearly with system size. Nevertheless, the formal scaling of finding all mRPA screened subspace interactions is still $\order{N^4}$ overall.

\section{Spin independence of mRPA interactions} 
\label{app:spin-independence}

In the case of a spin-conserving reference state where the irreducible RPA polarizability is the same for both spin-channels, and where the interaction kernel is spin-independent (e.g. in the absence of magnetic fields), the mRPA interaction resulting from the above procedure is also spin-independent. This allows for straightforward compatibility with methods to solve the effective Hamiltonian designed for use with the bare Coulomb interaction in `restricted' formulations, and furthermore allows for efficiency improvements in the construction of this effective interaction.

In such a case, $\mat{V}\mat{V}^T$ is the same for all spin-conserving transitions, so the Coulomb coupling between two spin-conserving excitations depends only on the spatial orbitals.
This combined with the fact that $\mat{\Delta}$ is spin-blocked means that when combining Eqs.~\ref{eq:Woodbury}~and~\ref{eq:final_NI_zero_mom_main} all terms containing the expansion of the Coulomb interaction over an auxiliary space intermediate are spin-independent.
The only spin-dependent contribution is thus the product of the two non-Coulombic terms in these equations, $\mat{\Delta}\mat{\Delta}^{-1} = \mat{I}$, and all other terms must be spin-independent.

When separated into blocks of single-particle excitations in each $(\alpha, \beta)$ spin channel the dRPA zeroth dd-moment must as such take the form
\begin{equation}
    \rpamom{0} = \left( 
    \begin{aligned}
        & \mat{I} & \mat{0} \\
        & \mat{0} & \mat{I} \\
    \end{aligned}
    \right)
    + 
    \left( 
    \begin{aligned}
        & \rpamom{0}_\text{s} & \rpamom{0}_\text{s} \\
        & \rpamom{0}_\text{s} & \rpamom{0}_\text{s} \\
    \end{aligned}
    \right),
\end{equation}
where $\rpamom{0}_\text{s}$ is a spin-independent, correlation-induced part of the zeroth dd-moment, depending only on the spatial component of the orbital basis. In this case, we expand the inverse of $\rpamom{0}$ in the spin-orbital basis via a Neumann series, as
\begin{widetext}
\begin{align}
        (\rpamom{0})^{-1} &= \sum_{n=0}^{\infty} (-1)^n \left( 
        \begin{aligned}
        & \rpamom{0}_\text{s} & \rpamom{0}_\text{s} \\
        & \rpamom{0}_\text{s} & \rpamom{0}_\text{s} \\
    \end{aligned}
    \right)^n \\
    &= \left( 
    \begin{aligned}
        & \mat{I} & \mat{0} \\
        & \mat{0} & \mat{I} \\
    \end{aligned}
    \right)
    - 
    \left( 
    \begin{aligned}
        & \rpamom{0}_\text{s} & \mat{0} \\
        & \mat{0} & \rpamom{0}_\text{s} \\
    \end{aligned}
    \right)
    \sum_{n=1}^{\infty} (-2)^{n-1} \left(
    \begin{aligned}
        & (\rpamom{0}_\text{s})^{n-1} & (\rpamom{0}_\text{s})^{n-1} \\
        & (\rpamom{0}_\text{s})^{n-1} & (\rpamom{0}_\text{s})^{n-1} \\
    \end{aligned}
    \right)\\
    &= \left( 
    \begin{aligned}
        & \mat{I} & \mat{0} \\
        & \mat{0} & \mat{I} \\
    \end{aligned}
    \right)
    -
    \left( 
    \begin{aligned}
        & \rpamom{0}_\text{s} & \mat{0} \\
        & \mat{0} & \rpamom{0}_\text{s} \\
    \end{aligned}
    \right)
        \left( 
    \begin{aligned}
        & (\mat{I} + 2\rpamom{0}_\text{s})^{-1}  & (\mat{I} + 2\rpamom{0}_\text{s})^{-1} \\
        & (\mat{I} + 2\rpamom{0}_\text{s})^{-1} & (\mat{I} + 2\rpamom{0}_\text{s})^{-1} \\
    \end{aligned}
    \right)
\end{align}
\end{widetext}

We can therefore write the inverse of the zeroth dd-moment compactly as
\begin{equation}
    {\rpamom{0}}^{-1} = \left( 
    \begin{aligned}
        & \mat{I} & \mat{0} \\
        & \mat{0} & \mat{I} \\
    \end{aligned}
    \right)
    - 
    \left( 
    \begin{aligned}
        & \mat{\kappa} & \mat{\kappa} \\
        & \mat{\kappa} & \mat{\kappa} \\
    \end{aligned}
    \right)
\end{equation}
with
\begin{equation}
    \mat{\kappa} = \rpamom{0}_\text{s} \left( 1 + 2 \rpamom{0}_\text{s} \right)^{-1}. \label{eq:mat_x}
\end{equation}
Insertion of this definition into Eq.~\ref{eq:inv_v} gives the spin-independent mRPA interactions for restricted references as
\begin{equation}
    v_\text{s} = \mat{\kappa} \mat{\Delta} \mat{\kappa} - \frac{1}{2} \left(\mat{\kappa} \mat{\Delta} + \mat{\Delta} \mat{\kappa}\right).
\end{equation}
Given that this only requires an inversion of a matrix in Eq.~\ref{eq:mat_x} which is half the size of the analogous spinned quantity in Eq.~\ref{eq:inv_v}, this can result in a more computationally efficient implementation with respect to model space size, where applicable. 

\section{RPA interacting bath orbitals} \label{app:RPAbath}

We seek to define a bath expansion for a quantum embedding to systematically expand the single particle space, in an optimally efficient way to describe the physics of the zeroth moment of the dd-response at the RPA level ($\rpamom{0}_{ia,jb}$). From a viewpoint of the static expectation values, this can be considered as the optimal single-particle bath expansion to span the fragment correlations of the two-body reduced density matrix at the RPA level. While the use of mRPA interactions in the cluster ensures that the subspace $\rpamom{0}$ is exactly reproduced at the level of RPA, this is not true when the cluster is solved at a higher level of theory, and thus a bath expansion can provide a route to systematically control the remaining error in this beyond-RPA description of the cluster. 

We define a bath expansion which will optimally minimize the error in the cluster $\rpamom{0}$ with {\em bare} interactions, ensuring that the bath spans these most relevant degrees of freedom at the level of the resulting solver, and in the process minimize the difference between the mRPA and bare interactions throughout the cluster for a given number of bath orbitals.
This approach is related to a similar (interacting) bath expansion which was introduced in Ref.~\cite{PhysRevX.12.011046} which optimally captured the one-body physics from a bare (M\o{}ller-Plesset) second-order perturbative level of theory. We expect that a screened perturbative bath expansion based on RPA to be more suitable for extended and polarizable systems, while noting that its construction relies on knowledge of the zeroth moment of the dd-response which is already required to construct the mRPA interactions in the resulting cluster, making the approach particularly straightforward and consistent in this framework. 

We start from a cluster formed from an arbitrary set of (generally local) `fragment' orbitals. These are augmented with the bath space from density matrix embedding theory (DMET), which ensures that the reference single determinant state is exactly spanned by a single state in the cluster, and that the mean-field density matrix of the fragment can be faithfully reproduced \cite{PhysRevLett.109.186404,doi:10.1021/ct301044e,sekaran2023unified,doi:10.1021/acs.jctc.6b00316,doi:10.1021/acs.jctc.9b00933,PhysRevX.12.011046}. We note that this construction also ensures the applicability of mRPA directly to this cluster. To obtain the RPA bath expansion orbitals to further augment this initial cluster, we form one-particle/hole pseudo-density matrices representing the correlated hole and particle fluctuations at the level of RPA which couple this initial cluster to its wider environment,
\begin{align}
    \gamma^{\mathbf{x}}_{ij} &= - \sum\limits_{\tilde{a}} (\rpamom{0}_{i\tilde{a}j\tilde{a}} - \delta_{ij}\delta_{\tilde{a}\tilde{a}})\label{eq:RPA_oo_rdm}\\
    \gamma^{\mathbf{x}}_{ab} &= - \sum\limits_{\tilde{i}} (\rpamom{0}_{\tilde{i}a\tilde{i}b} - \delta_{ab}\delta_{\tilde{i}\tilde{i}})\label{eq:RPA_vv_rdm}.
\end{align}
Here, $\tilde{i}$, ($\tilde{a}$) indices denote the local hole (particle) states associated with the initial cluster, $\mathbf{x}$, which can be efficiently implemented via projection operators, while other indices represent environmental hole (particle) orbitals in the complementary space~\cite{PhysRevX.12.011046}.
Note that all indices represent spin-orbitals, which in spin colinear systems will require that all indices have the same spin label.
As such, from this point on all indices will be assumed to represent the same spin channel.

The eigenvectors of these matrices represent orbitals of a bath expansion of the correlated environment in occupied and unoccupied states, with their corresponding eigenvalues quantifying their importance in describing the full RPA physics of $\rpamom{0}$ in the cluster. These eigenvalues can then be used as a systematic truncation via a threshold in order to define the size of the bath space, analogous to the approach in Ref.~\onlinecite{PhysRevX.12.011046}. The mRPA (or bare) interactions can then be found in this subspace for the high-level solver in a quantum embedding framework.
The required $\rpamom{0}_{i\tilde{a}j\tilde{a}}$ components can be calculated in $\order{N^4}$ computational cost 
based on the approach in App.~\ref{app:compute}, with additional techniques (not pursued here) expected to be able to reduce this to $\order{N^3}$ via restriction of the full excitation space in the construction of $\rpamom{0}_{i\tilde{a}j\tilde{a}}$, or working with a doubly factorized bare interaction \cite{10.1063/5.0143291}.

To provide some further insight into the nature of these bath orbitals, we can explicitly consider the relation between the two-body density matrix, $\Gamma_{pqrs}$, and the zeroth moment of the four-point dd-response in an arbitrary basis as an expansion over all neutral excitations,
\begin{equation}
    \chi^{(0)}_{pqrs} = \sum_{\nu\neq 0} \bra{0} \hat{c}_q^\dagger \hat{c}_p \ket{\nu} \bra{\nu} \hat{c}_s^\dagger \hat{c}_r \ket{0}
    = \Gamma_{pqsr} - \gamma_{pq} \gamma_{rs} + \delta_{qs} \gamma_{pr},
    \label{eq:relation_eta0_2rdm}
\end{equation}
where $\chi^{(0)}_{pqrs}$ is the zeroth moment of the dd-response, which is related to the $\rpamom{0}$ quantity central to this work via summation over the excitations and de-excitation manifold, 
\begin{equation}
    \rpamom{0}_{iajb} = \chi^{(0)}_{iajb} + \chi^{(0)}_{iabj} + \chi^{(0)}_{aijb} + \chi^{(0)}_{aibj}.
\end{equation}
We find that the RPA pseudo-density matrices for each cluster $\mathbf{x}$ used to define the RPA bath expansion in Eqs.~\ref{eq:RPA_oo_rdm}-\ref{eq:RPA_vv_rdm} are related to a hypothetical RPA one-body density matrix, $\gamma$, as
\begin{align}
    \gamma^{\mathbf{x}}_{ij} &= \sum_{\tilde{a}}\left[ 
    (\delta_{ij} - \gamma_{ij}) (\delta_{\tilde{a}\tilde{a}} - \gamma_{\tilde{a}\tilde{a}})
    + \gamma_{\tilde{a}\tilde{a}} \gamma_{ij} + 4 \gamma_{i\tilde{a}}\gamma_{j\tilde{a}}
    \right]  \\
    \gamma^{\mathbf{x}}_{ab} &= 
    \sum_{\tilde{i}}\left[ 
    (\delta_{\tilde{i}\tilde{i}} - \gamma_{\tilde{i}\tilde{i}}) (\delta_{ab} - \gamma_{ab})
    + \gamma_{\tilde{i}\tilde{i}} \gamma_{ab} + 4 \gamma_{\tilde{i}a}\gamma_{\tilde{i}b}
    \right].
\end{align}
While this does not guarantee positive definiteness of the quantities in Eqs.~\ref{eq:RPA_oo_rdm}-\ref{eq:RPA_vv_rdm}, since the RPA density matrices are not well-defined~\cite{kosov2017restoring} (hence the use of {\em pseudo-}density matrix), it gives some physical justification for such a property.
It also demonstrates that the expansion captures the one-body coupling to the environment, despite originating from a purely two-body correlated quantity. 

\section{Non-local and energetic expectation values in graphene} \label{app:graphene}

\begin{figure}
    \centering
    \includegraphics[width=\linewidth]{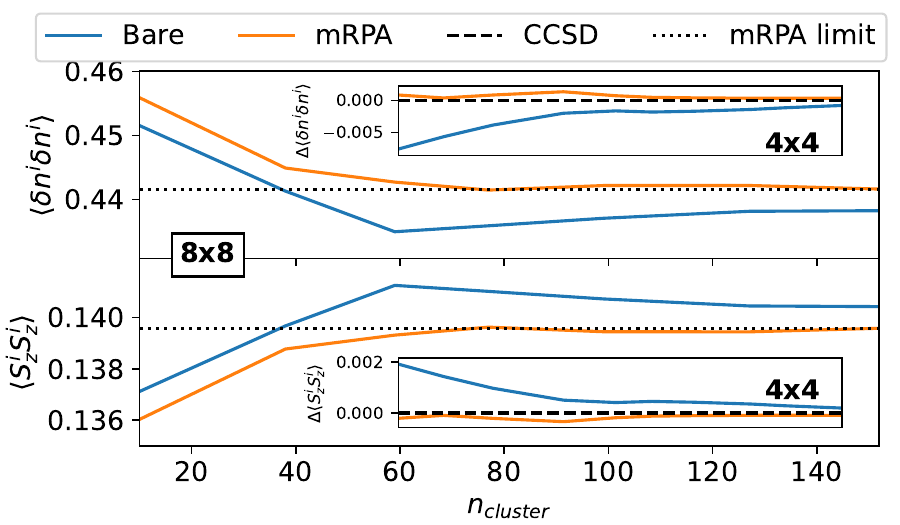}
    \caption{Analogous results to Fig.~\ref{fig:graphene-main} of the main text, but all bath orbitals and mRPA interactions are constructed on a Kohn--Sham PBE rather than Hartree--Fock reference state. Convergence of the CCSD local $2p_z$-orbital charge (top) and spin (bottom) two-body fluctuations in $8\times8$ Graphene with increasing size of embedded cluster, for both bare Coulomb and mRPA cluster interactions. Inset: Error in these two-body correlators in $4\times4$ $\textbf{k}-$point meshes, where comparison to exact CCSD (on the PBE reference) is possible.}
    \label{fig:PBE-ref}
\end{figure}

In this section, we expand on the results of mRPA and quantum embedding calculations on the graphene system, considering the effect of different reference states, as well as the impact of the mRPA screening on non-local expectation values. The results of the main text were performed for a $4 \times 4$ and $8 \times 8$ $\Gamma$-centered $\mathbf{k}$-point mesh, in a cc-pVDZ basis with a restricted Hartree--Fock reference state, via the {\tt Vayesta} quantum embedding package, which interfaces with the {\tt pyscf} codebase~\cite{10.1063/5.0006074,https://doi.org/10.1002/wcms.1340,doi:10.1021/acs.jctc.9b00933,doi:10.1021/acs.jctc.7b00049}. However, no double-counting results from the use of Kohn--Sham references in order to define the orbitals and their energies in the irreducible polarizability instead of Hartree--Fock. There may be advantages in this in terms of convergence of the mean-field solution in metallic systems and accuracy of the resulting mRPA interactions in extended systems. However, too much dependence on this starting reference would also be a cause of concern and an indication of also inheriting the empiricism in the choice of mean-field theory.

In Fig.~\ref{fig:PBE-ref} we present analogous calculations to those of Fig.~\ref{fig:graphene-main} in the main text, showing the local charge and spin correlators with bare and mRPA interactions in the quantum embedding of increasing cluster sizes, but now on a Kohn--Sham reference state with the PBE functional. This can change both the construction of the RPA bath orbitals of App.~\ref{app:RPAbath} defining the single-particle space of the cluster, as well as the mRPA interactions which are used in its subsequent solution. While the mean-field gap is quite different to the Hartree--Fock reference, the convergence of the CCSD local correlators from the clusters for both mRPA and bare interactions is almost identical. This is reassuring that it is genuinely describing the effect of the post-mean-field physics, and inherits the largely reference-independent property of the CCSD solver in the full system limit. Future work will consider a self-consistent approach to the mRPA interactions, formally removing the mean-field dependence.

\begin{figure}
    \centering
    \includegraphics[width=\linewidth]{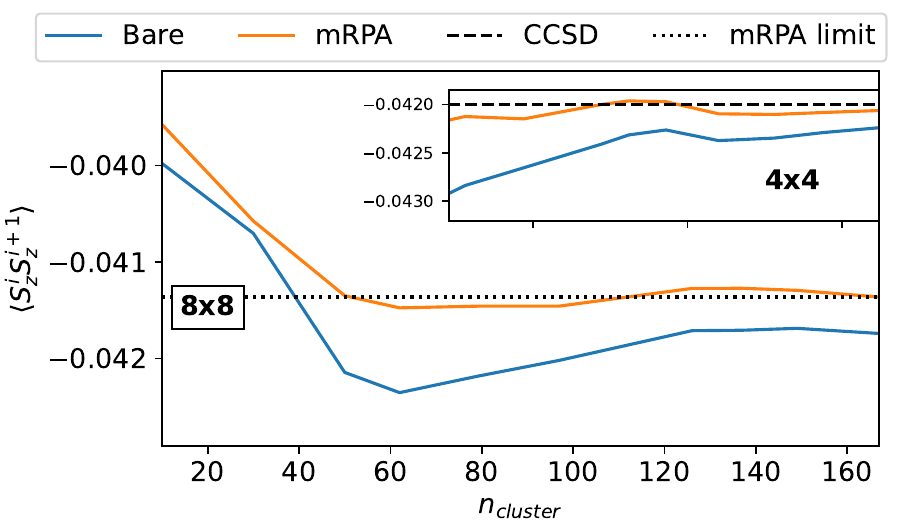}
    \caption{Convergence of the CCSD nearest-neighbor $2p_z$-orbital spin correlated fluctuations in $8\times8$ Graphene with increasing size of embedded cluster, for both bare Coulomb and mRPA cluster interactions. Inset: Error in this two-body correlators in $4\times4$ $\textbf{k}-$point mesh, where comparison to exact CCSD is possible.}
    \label{fig:non-local-spin}
\end{figure}

We also consider the accuracy of {\em non-local} expectation values in a quantum embedding framework with mRPA screened interactions, where the operators span more than one fragment (impurity). One approach is to simply enlarge the fragment space to encompass the range of the operator of interest. However, this is inefficient and will not be suitable for full system expectation values such as total energies. Reference~\onlinecite{doi:10.1021/acs.jctc.2c01063} considered approaches to couple the solutions of different embedded quantum clusters for non-local expectation values. We therefore follow this prescription to compute atomic $2p_z$ spin correlators to quantify the fluctuations {\em between} two nearest-neighbour atoms (labelled $A$ and $B$), as $\langle {\hat S_z^A} {\hat S_z^B \rangle}$ in Fig.~\ref{fig:non-local-spin}, via construction of the two-body reduced density matrix (denoted $\Gamma[(\gamma, K^*)[\Psi^{\mathbf{x}}]]$ in Ref.~\onlinecite{doi:10.1021/acs.jctc.2c01063}). We highlight that the atomic fragmentation means that these correlators span different embedded cluster solutions. 

These results again show the impact of the mRPA screening in the cluster solutions for this system, even for non-local expectation values which are not explicitly conserved in the mRPA construction. Plasmonic modes over the longest length-scales are folded in to the local interactions, reducing the nearest-neighbour magnetic fluctuations, and enabling high-accuracy and systematic convergence compared to bare interacting clusters.

When extending this analysis to total energies, some more care is required. The magnetic correlators considered extend beyond the length of any single fragment, but the union of all fragments do span the considered operators. This is not always the case however, and in our embedding protocol where fragments are intrinsic atomic orbitals of minimal size, the space of high-energy unoccupied states in large basis sets is not contained in any fragment. While the mRPA interactions will relax and improve the description of the correlated low-energy fragment spaces when solved in isolation, they will not account for the incomplete trace in the evaluation of non-local expectation values over these high energy degrees of freedom not spanned by any of the solved subspaces. 

When considering total energies with variational solvers of fragmented (yet incomplete) subspaces, this incompleteness in the space spanned by the Hamiltonian in the union of the subspaces will serve to increase the total energy of the fragmented system. Similarly, the use of mRPA interactions in the subspaces will serve to also increase the energy of the individual subspace descriptions. Despite the fact that the description of these subspaces with mRPA interactions is closer to the projection of the full system description into this subspace, the use of mRPA interactions will therefore necessarily serve to {\em increase} the total energy error. This is shown in Fig.~\ref{fig:energy_graphene}, where the correlation energy error of graphene increases with mRPA interactions. This energy over the embedded fragments is defined via the `projected' energy estimate, introduced in Ref.~\onlinecite{PhysRevX.12.011046}, and expanded on in Ref.~\onlinecite{doi:10.1021/acs.jctc.2c01063}. It should be noted that this argument of energy error increase with mRPA interactions is not necessarily true when considering energy differences.

\begin{figure}
    \centering
    \includegraphics[width=\linewidth]{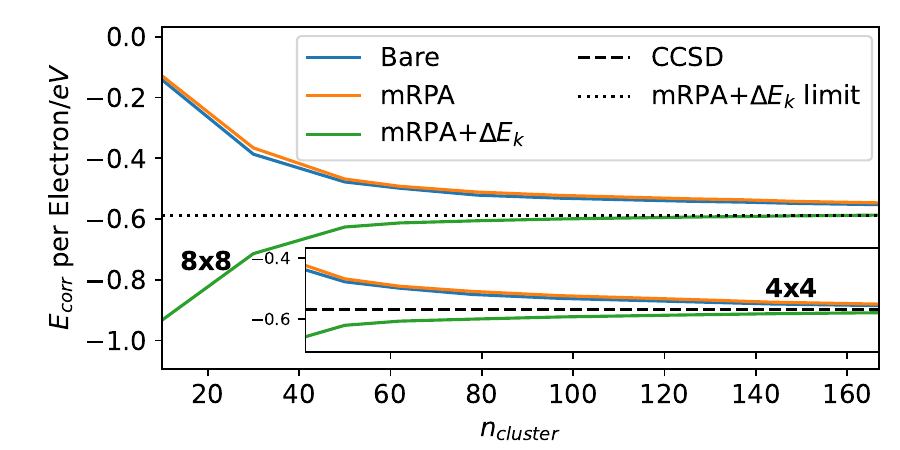}
    \caption{Convergence of the fragmented CCSD correlation energy in $8\times8$ Graphene with increasing size of embedded cluster, for bare Coulomb interactions and mRPA cluster interactions with and without the $\Delta E_k$ RPA energy correction. Inset: Equivalent error in the correlation energy in a $4\times4$ $\textbf{k}-$point mesh, where comparison to full system CCSD is possible. Calculations were based upon a RHF reference.}
    \label{fig:energy_graphene}
\end{figure}

We therefore seek a way to improve the convergence of energy estimators in keeping with the spirit of mRPA, to include the energetic contributions from high-energy and long-range contributions that are not contained within any subspace description. This would therefore directly include the energetic effect of e.g. plasmon modes, rather than just the effect of their coupling on the low-energy descriptions. The brute-force approach via an (interacting) bath expansion can be seen in Fig.~\ref{fig:energy_graphene} to be poorly convergent, while previous DMET approaches have considered direct enlargement of fragment spaces to include these energetic contributions, but substantially increases the cost of the subspace solutions~\cite{doi:10.1021/acs.jctc.9b00933}. 

As an alternative approach, we formulate a full-system energetic correction to these incomplete fragmented subspace expectation values via the mRPA construction.
We begin by noting that we can construct a global two-body density matrix contribution from the RPA via the relation given in Eq.~\ref{eq:relation_eta0_2rdm}.
$\chi^{(0)}_{pqrs}$ then relates to the cumulant (or connected) component of the two-body density matrix, $K_{pqrs}$, as 
\begin{equation}
    K_{pqrs} = \chi^{(0)}_{pqsr} - \gamma_{ps} ( \delta_{rq} - \gamma_{rq}). \label{eq:cumulant_dd_relation}
\end{equation}
This represents the correlated, two-body correlations not decomposable as a combination of one-body correlations.
The contribution of these irreducible two-body fluctuations to the total energy is given by
\begin{align}
    E_k &= \frac{1}{2} \sum_{pqrs} (pq|rs) K_{pqrs} \\
    &= \frac{1}{2} \sum_{pqrs} (pq|rs) \left(\chi^{(0)}_{pqsr} - \gamma_{ps} ( \delta_{rq} - \gamma_{rq})\right).
\end{align}
Under the RPA description of $ph$ (de-)excitations, the first term naturally becomes
\begin{align}
    \sum_{pqrs} (pq|rs) \chi^{(0)}_{pqsr} = \sum_{\substack{ij\in \text{occ}\\ab\in \text{vir}}} (ia|jb)\rpamom{0}_{iajb}
\end{align}
while approximating $\gamma$ by its mean-field value in the remaining terms gives
\begin{align}
    \sum_{pqrs} (pq|rs) \gamma_{ps} ( \delta_{rq} - \gamma_{rq}) &\approx \sum_{pqrs} \delta_{ps}^{p \in \text{occ}} \delta_{rq}^{r \in \text{vir}} (pq|rs) \nonumber\\
    &= \sum_{\substack{i\in \text{occ}\\a\in \text{vir}}} (ia|ia).
\end{align}

We therefore obtain an overall correlation energy contribution from these irreducible two-body fluctuations as
\begin{equation}
    E_k = \frac{1}{2} \text{Tr} \left[\left(\mat{\eta}^{(0)} - \mat{I} \right)\mat{v}\right]. \label{eq:Ek}
\end{equation}

To use this as a correction to our correlated energy estimate obtained via quantum embedding of our incomplete fragmented wavefunction, we must remove local contributions from each cluster to avoid double counting.
This can be achieved simply in our quantum embedding approach by computing a contribution to these two-body energetic fluctuations in each subspace, by projecting one index into the fragment space, and the remaining indices restricted to the subspace of consideration~\cite{doi:10.1021/acs.jctc.2c01063}. This provides a decomposition of the total RPA two-body energy for a cluster $\mathbf{x}$ as
\begin{equation}
    E_k^\mathbf{x} = \frac{1}{2} \text{Tr}\left[
    \mat{P}^\mathbf{x}
    \left(\mat{\eta}^{(0)} - \mat{I} \right)\mat{v}\right],
\end{equation}
where $\mat{P}^\mathbf{x}$ is defined as symmetrically projecting one of the occupied indices of the full-system $\rpamom{0}$ into the fragment space, with all other indices projected into the cluster space of $\mathbf{x}$ (as in Eq.~25 of \cite{doi:10.1021/acs.jctc.2c01063}). 

The final non-local energy correction is therefore obtained by deducting the local subspace contributions from the global expectation, giving the result
\begin{equation}
    \Delta E_k = E_k - \sum_\mathbf{x} E_k^\mathbf{x}.
\end{equation}
This is equivalent to only including contributions in Eq.~\ref{eq:Ek} where all indices are not contained within any of their cluster space.  While this choice of decomposition into subspace contributions to avoid double-counting is specific for our approach for computing expectation values, the general approach could be easily adapted for other embedding or subspace methodologies in order to define a non-local RPA energy correction.

Application of this RPA energetic two-body correction for the incompleteness of the cluster spaces in the atomic IAO quantum embedding of graphene is shown in Fig.~\ref{fig:energy_graphene}. This is found to substantially improve the rate of convergence with respect to the size of the explicit CCSD clusters towards the full CCSD limit, with the energy correction accurately compensating for the long-range bubble contributions not spanned by the clusters. Other energy contributions are expected to decay more rapidly, and therefore more easily described by the increasing local cluster spaces.
\fi

\end{document}